\documentclass[twocolumn]{aastex631}

\usepackage{amsmath}
\usepackage{verbatim}

\newcommand{\thisevent}{OGLE-2016-BLG-1093}
\newcommand{\Spitzer}{{\em Spitzer}}

\def\au{{\rm AU}} 

\newcommand{\pivec}{\mbox{\boldmath $\pi$}}

\newcommand{\pieN}{\mbox{$\pi_{{\rm E},N}$}}
\newcommand{\pieE}{\mbox{$\pi_{{\rm E},E}$}}
\newcommand{\Apie}{\mbox{$|\pi_{{\rm E}}|$}}

\lefthead{} 
\righthead{}

\begin{document}

\title{
OGLE-2016-BLG-1093Lb: A Sub-Jupiter-mass \Spitzer\ Planet Located in Galactic Bulge
}

\author{In-Gu Shin} 
\affiliation{Korea Astronomy and Space Science Institute, Daejon 34055, Republic of Korea}
\author{Jennifer C. Yee}
\affiliation{Center for Astrophysics $|$ Harvard \& Smithsonian 60 Garden St., Cambridge, MA 02138, USA}
\author{Kyu-Ha Hwang}
\affiliation{Korea Astronomy and Space Science Institute, Daejon 34055, Republic of Korea}
\author{Andrew Gould}
\affiliation{Max Planck Institute for Astronomy, K\"onigstuhl 17, D-69117 Heidelberg, Germany}
\affiliation{Department of Astronomy, The Ohio State University, 140 W. 18th Ave., Columbus, OH 43210, USA}
\author{Andrzej Udalski}
\affiliation{Astronomical Observatory, University of Warsaw, Al.~Ujazdowskie 4, 00-478 Warszawa, Poland}
\author{Ian A. Bond}
\affiliation{Institute of Information and Mathematical Sciences, Massey University, Private Bag 102-904, North Shore Mail Centre, Auckland, New Zealand}
\collaboration{7}{(Leading authors),}
%
\author{Michael D. Albrow} 
\affiliation{University of Canterbury, Department of Physics and Astronomy, Private Bag 4800, Christchurch 8020, New Zealand}
\author{Sun-Ju Chung}
\affiliation{Korea Astronomy and Space Science Institute, Daejon 34055, Republic of Korea}
\affiliation{University of Science and Technology, Korea, (UST), 217 Gajeong-ro, Yuseong-gu, Daejeon 34113, Republic of Korea}
\author{Cheongho Han}
\affiliation{Department of Physics, Chungbuk National University, Cheongju 28644, Republic of Korea}
\author{Youn Kil Jung}
\affiliation{Korea Astronomy and Space Science Institute, Daejon 34055, Republic of Korea}
\affiliation{University of Science and Technology, Korea, (UST), 217 Gajeong-ro, Yuseong-gu, Daejeon 34113, Republic of Korea}
\author{Hyoun Woo Kim}
\affiliation{Korea Astronomy and Space Science Institute, Daejon 34055, Republic of Korea}
\author{Yoon-Hyun Ryu}
\affiliation{Korea Astronomy and Space Science Institute, Daejon 34055, Republic of Korea}
\author{Yossi Shvartzvald}
\affiliation{Department of Particle Physics and Astrophysics, Weizmann Institute of Science, Rehovot 76100, Israel}
\author{Weicheng Zang}
\affiliation{Department of Astronomy and Tsinghua Centre for Astrophysics, Tsinghua University, Beijing 100084, China}
%
\author{Sang-Mok Cha}
\affiliation{Korea Astronomy and Space Science Institute, Daejon 34055, Republic of Korea}
\affiliation{School of Space Research, Kyung Hee University, Yongin, Kyeonggi 17104, Republic of Korea}
\author{Dong-Jin Kim}
\affiliation{Korea Astronomy and Space Science Institute, Daejon 34055, Republic of Korea}
\author{Seung-Lee Kim} 
\affiliation{Korea Astronomy and Space Science Institute, Daejon 34055, Republic of Korea}
\author{Chung-Uk Lee}
\affiliation{Korea Astronomy and Space Science Institute, Daejon 34055, Republic of Korea}
\author{Dong-Joo Lee}
\affiliation{Korea Astronomy and Space Science Institute, Daejon 34055, Republic of Korea}
\author{Yongseok Lee}
\affiliation{Korea Astronomy and Space Science Institute, Daejon 34055, Republic of Korea}
\affiliation{School of Space Research, Kyung Hee University, Yongin, Kyeonggi 17104, Republic of Korea}
\author{Byeong-Gon Park}
\affiliation{Korea Astronomy and Space Science Institute, Daejon 34055, Republic of Korea}
\affiliation{University of Science and Technology, Korea, (UST), 217 Gajeong-ro, Yuseong-gu, Daejeon 34113, Republic of Korea}
\author{Richard W. Pogge}
\affiliation{Department of Astronomy, The Ohio State University, 140 W. 18th Ave., Columbus, OH 43210, USA}
\collaboration{17}{(The KMTNet Collaboration),}
%
\author{Przemek Mr{\'o}z}
\affiliation{Astronomical Observatory, University of Warsaw, Al.~Ujazdowskie 4, 00-478 Warszawa, Poland}
\author{Micha{\l} K. Szyma{\'n}ski}
\affiliation{Astronomical Observatory, University of Warsaw, Al.~Ujazdowskie 4, 00-478 Warszawa, Poland}
\author{Jan Skowron}
\affiliation{Astronomical Observatory, University of Warsaw, Al.~Ujazdowskie 4, 00-478 Warszawa, Poland}
\author{Radek Poleski}
\affiliation{Astronomical Observatory, University of Warsaw, Al.~Ujazdowskie 4, 00-478 Warszawa, Poland}
\author{Igor Soszy{\'n}ski}
\affiliation{Astronomical Observatory, University of Warsaw, Al.~Ujazdowskie 4, 00-478 Warszawa, Poland}
\author{Pawe{\l} Pietrukowicz}
\affiliation{Astronomical Observatory, University of Warsaw, Al.~Ujazdowskie 4, 00-478 Warszawa, Poland}
\author{Szymon Koz{\l}owski}
\affiliation{Astronomical Observatory, University of Warsaw, Al.~Ujazdowskie 4, 00-478 Warszawa, Poland}
\author{Krzysztof Ulaczyk}
\affiliation{Department of Physics, University of Warwick, Gibbet Hill Road, Coventry, CV4 7AL, UK}
\collaboration{9}{(The OGLE Collaboration)} 
%
\author{Charles A. Beichman}
\affiliation{IPAC, Mail Code 100-22, Caltech, 1200 E. California Blvd., Pasadena, CA 91125, USA}
\author{Geoffery Bryden}
\affiliation{Jet Propulsion Laboratory, California Institute of Technology, 4800 Oak Grove Drive, Pasadena, CA 91109, USA}
\author{Sebastiano Calchi Novati}
\affiliation{IPAC, Mail Code 100-22, Caltech, 1200 E. California Blvd., Pasadena, CA 91125, USA}
\author{Sean Carey}
\affiliation{IPAC, Mail Code 100-22, Caltech, 1200 E. California Blvd., Pasadena, CA 91125, USA}
\author{B. Scott Gaudi}
\affiliation{Department of Astronomy, Ohio State University, 140 W. 18th Ave., Columbus, OH  43210, USA}
\author{Calen B. Henderson}
\affiliation{IPAC, Mail Code 100-22, Caltech, 1200 E. California Blvd., Pasadena, CA 91125, USA}
\author{Wei Zhu}
\affiliation{Department of Astronomy, Tsinghua University, Beijing 100084, China}
\collaboration{8}{(the \Spitzer\ team)}
%
\author{Fumio Abe}
\affiliation{Institute for Space-Earth Environmental Research, Nagoya University, Nagoya 464-8601, Japan}
\author{Richard K. Barry}
\affiliation{Astrophysics Science Division, NASA/Goddard Space Flight Center, Greenbelt, MD20771, USA}
\author{David P. Bennett}
\affiliation{Laboratory for Exoplanets and Stellar Astrophysics, NASA / Goddard Space Flight Center, Greenbelt, MD 20771, USA}
\affiliation{Department of Astronomy, University of Maryland, College Park, MD 20742, USA}
\author{Aparna Bhattacharya}
\affiliation{Laboratory for Exoplanets and Stellar Astrophysics, NASA / Goddard Space Flight Center, Greenbelt, MD 20771, USA}
\affiliation{Department of Astronomy, University of Maryland, College Park, MD 20742, USA}
\author{Hirosane Fujii}
\affiliation{Institute for Space-Earth Environmental Research, Nagoya University, Nagoya 464-8601, Japan}
\author{Akihiko Fukui}
\affiliation{Department of Earth and Planetary Science, Graduate School of Science, The University of Tokyo, 7-3-1 Hongo, Bunkyo-ku, Tokyo 113-0033, Japan}
\affiliation{Instituto de Astrof\'isica de Canarias, V\'ia L\'actea s/n, E-38205 La Laguna, Tenerife, Spain}
\author{Yuki Hirao}
\affiliation{Depertment of Earth and Space Science, Graduate School of Science, Osaka University, 1-1 Machikaneyama, Toyonaka, Osaka 560-0043, Japan}
\affiliation{Laboratory for Exoplanets and Stellar Astrophysics, NASA / Goddard Space Flight Center, Greenbelt, MD 20771, USA}
\affiliation{Department of Astronomy, University of Maryland, College Park, MD 20742, USA}
\author{Yoshitaka Itow}
\affiliation{Institute for Space-Earth Environmental Research, Nagoya University, Nagoya 464-8601, Japan}
\author{Rintaro Kirikawa}
\affiliation{Department of Earth and Space Science, Graduate School of Science, Osaka University, Toyonaka, Osaka 560-0043, Japan}
\author{Naoki Koshimoto}
\affiliation{Department of Astronomy, Graduate School of Science, The University of Tokyo, 7-3-1 Hongo, Bunkyo-ku, Tokyo 113-0033, Japan}
\author{Iona Kondo}
\affiliation{Depertment of Earth and Space Science, Graduate School of Science, Osaka University, 1-1 Machikaneyama, Toyonaka, Osaka 560-0043, Japan}
\author{Yutaka Matsubara}
\affiliation{Institute for Space-Earth Environmental Research, Nagoya University, Nagoya 464-8601, Japan}
\author{Sho Matsumoto}
\affiliation{Department of Earth and Space Science, Graduate School of Science, Osaka University, Toyonaka, Osaka 560-0043, Japan}
\author{Shota Miyazaki}
\affiliation{Depertment of Earth and Space Science, Graduate School of Science, Osaka University, 1-1 Machikaneyama, Toyonaka, Osaka 560-0043, Japan}
\author{Yasushi Muraki}
\affiliation{Institute for Space-Earth Environmental Research, Nagoya University, Nagoya 464-8601, Japan}
\author{Greg Olmschenk}
\affiliation{Code 667, NASA Goddard Space Flight Center, Greenbelt, MD 20771, USA}
\author{Arisa Okamura}
\affiliation{Department of Earth and Space Science, Graduate School of Science, Osaka University, Toyonaka, Osaka 560-0043, Japan}
\author{Cl\'ement Ranc}
\affiliation{Sorbonne Universit\'e, CNRS, UMR 7095, Institut d'Astrophysique de Paris, 98 bis bd Arago, 75014 Paris, France}
\author{Nicholas J. Rattenbury}
\affiliation{Department of Physics, University of Auckland, Private Bag 92019, Auckland, New Zealand}
\author{Yuki Satoh}
\affiliation{Department of Earth and Space Science, Graduate School of Science, Osaka University, Toyonaka, Osaka 560-0043, Japan}
\author{Stela Ishitani Silva}
\affiliation{Department of Physics, The Catholic University of America, Washington, DC 20064, USA}
\affiliation{Code 667, NASA Goddard Space Flight Center, Greenbelt, MD 20771, USA}
\author{Takahiro Sumi}
\affiliation{Depertment of Earth and Space Science, Graduate School of Science, Osaka University, 1-1 Machikaneyama, Toyonaka, Osaka 560-0043, Japan}
\author{Daisuke Suzuki}
\affiliation{Depertment of Earth and Space Science, Graduate School of Science, Osaka University, 1-1 Machikaneyama, Toyonaka, Osaka 560-0043, Japan}
\author{Taiga Toda}
\affiliation{Department of Earth and Space Science, Graduate School of Science, Osaka University, Toyonaka, Osaka 560-0043, Japan}
\author{Paul J. Tristram}
\affiliation{University of Canterbury Mt. John Observatory, P.O. Box 56, Lake Tekapo 8770, New Zealand}
\author{Aikaterini Vandorou}
\affiliation{Code 667, NASA Goddard Space Flight Center, Greenbelt, MD 20771, USA}
\affiliation{Department of Astronomy, University of Maryland, College Park, MD 20742, USA}
\author{Hibiki Yama}
\affiliation{Department of Earth and Space Science, Graduate School of Science, Osaka University, Toyonaka, Osaka 560-0043, Japan}
\collaboration{28}{(the MOA Collaboration)}

\bigskip\bigskip
%

\begin{abstract}
\thisevent\ is a planetary microlensing event that is part of the statistical \Spitzer\ microlens parallax sample. The precise measurement of the microlens parallax effect for this event, combined with the measurement of finite source effects, leads to a direct measurement of the lens masses and system distance: $M_{\rm host} = 0.38$--$0.57\, M_{\odot}$, $m_p = 0.59$--$0.87\, M_{\rm Jup}$, and the system is located at the Galactic bulge ($D_L \sim 8.1$ kpc). Because this was a high-magnification event, we are also able to empirically show that the ``cheap-space parallax" concept \citep{gould12} produces well-constrained (and consistent) results for $\Apie$. This demonstrates that this concept can be extended to many two-body lenses. Finally, we briefly explore systematics in the \Spitzer\ light curve in this event and show that their potential impact is strongly mitigated by the color-constraint.
\end{abstract}

\keywords{Gravitational microlensing (672) --- Microlensing parallax (2144) --- Gravitational microlensing exoplanet detection (2147)}

\section{Introduction}

The effect of the Bulge environment on planet formation has yet to be determined. A few early studies, such as \citet{Gonzalez01} and \citet{Lineweaver04}, investigated how various properties that vary throughout the Galaxy, such as metallicity and supernova rate, might impact planet formation. These issues were subsequently revisited in \citet{Gowanlock11}. Later, \citet{thompson13} suggested the ambient temperature of the Bulge could inhibit the formation of ices, and thus, of giant planets. Because of its ability to find planets in both the Disk and the Bulge of the Galaxy, microlensing is the best technique for directly measuring the frequency of Bulge planets.

Two statistical studies have attempted to address the relative frequency of Disk and Bulge planets. \citet{Penny16} compared the distances (some measured and some estimated with a Bayesian analysis) of $31$ known microlensing planets with the expected distribution from a Galactic model for a range of relative Disk/Bulge planet frequencies. Their limit on the relative planet frequency suggests fewer or no planets in the Bulge relative to the Disk. Then, \citet{Koshimoto21} published an analysis comparing the observed lens-source relative proper motion, $\mu_{\rm rel}$, and Einstein timescale, $t_{\rm E}$, distributions with the expectations from a Galactic model. They find the distribution is consistent with no dependence on Galactocentric radius, but with large uncertainties. Part of the reason their result is so imprecise is that $\mu_{\rm rel}$ and $t_{\rm E}$ only provide a mass-distance {\it relation} for each object in the sample.

By contrast to these previous two studies, the \Spitzer\ microlensing parallax program was undertaken to directly measure distances to planets in order to infer the relative frequency of planets in the Disk and the Bulge \citep{CalchiNovati15,yee15,zhu17}. In this paper, we present the analysis of \thisevent, the eighth planet in the statistical sample of \Spitzer\ planets. We begin with an overview of the observations in Section \ref{sec:data}. The analysis of the ground-based light curve is presented in Section \ref{sec:gb}. In Section, \ref{sec:spz_par}, we fit the \Spitzer\ data for the satellite microlens parallax effect. We discuss various tests of the \Spitzer\ parallax and low-level systematics in the \Spitzer\ light curve in Section \ref{sec:Spz_data}. We derive the physical properties of the lens in Section \ref{sec:cmd} and verify membership in the \Spitzer\ sample in Section \ref{sec:spzmember}. Finally, we give our conclusions in Section \ref{sec:conclu}.

\begin{figure*}[htb!]
\epsscale{1.10}
\plotone{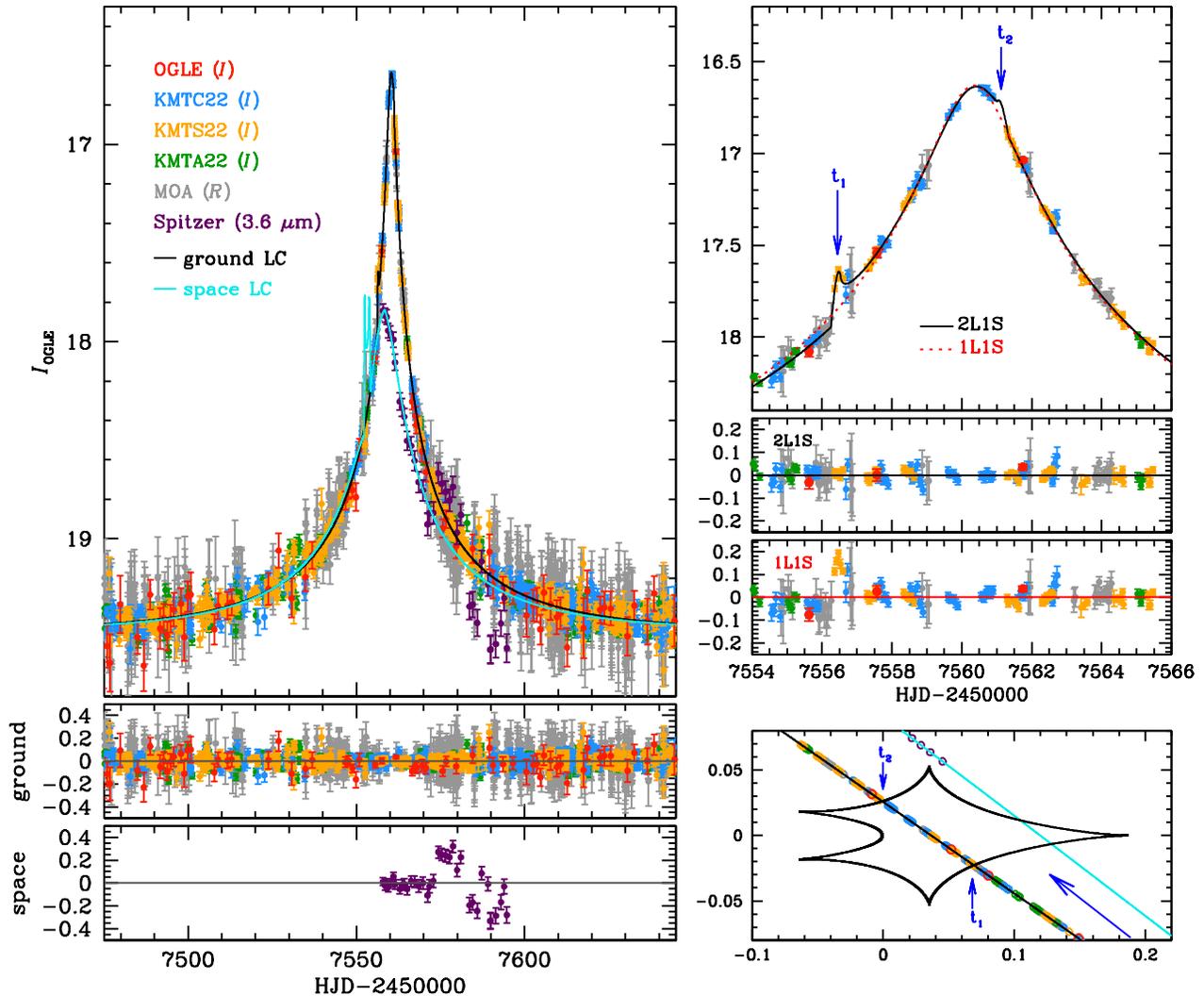}
\caption{
Light curve with the best-fit model (see Table \ref{table:space}). The left panels show the full view of the light curve from both 
the ground and \Spitzer. The upper-right panels show just the ground-based data for the the anomaly 
part of the light curve (caustic crossings at $t_1$ and $t_2$), which is induced by the planetary 
system. For comparison, we present the 1L1S model light curve (dashed red line) with residuals. 
The lower-right panel presents the caustic geometry of the event. 
\label{fig:lc}}
\end{figure*}

{\section{Observations}
\label{sec:data}}

The microlensing event occurred on a background star (i.e., source) located at $(\alpha, \delta)=(17^{h} 56^{m} 01^{s}.03,\, -32^{\circ} 42{'} 48{''}.5)$ in equatorial coordinates, which corresponds to $(\ell,b)=(-2^{\circ}.108,\, -3^{\circ}.848)$ in Galactic coordinates. This event was first announced by the Early Warning System \citep[EWS;][]{udalski94, udalski03} of the Optical Gravitational Lensing Experiment survey \citep[OGLE-IV;][]{udalski15} on 2016 June 11. Thus, this event is designated \thisevent. The event was observed using the $1.3\,m$ Warsaw telescope ($1.4\,{\rm deg}^{2}$ science camera) located at Las Campanas Observatory in Chile.  

Two other microlensing surveys independently discovered this event. The Microlensing Observations in Astrophysics \citep[MOA;][]{bond01, sumi03} detected this event on June $24$, $2016$, using the $1.8\,m$ telescope located at the University of Canterbury Mount John Observatory in New Zealand. The Korea Microlensing Telescope Network \citep[KMTNet;][]{kim16} also detected this event using a telescope network consisting of three identical $1.6\,m$ telescopes ($4\,{\rm deg}^{2}$ science cameras) located at the Cerro Tololo Inter-American Observatory in Chile (KMTC), the South African Astronomical Observatory in South Africa (KMTS), and the Siding Spring Observatory in Australia (KMTA). As a result, the light curve of \thisevent\ is well covered by five ground-based observatories (see Figure \ref{fig:lc}).

In addition, this event was chosen as a target of the {\it Spitzer Microlensing Campaign} (see \citealt{yee15} for the details of target selection criteria) and observed using the {\it Spitzer Space Telescope} with the $3.6\, {\mu m}$ channel of the IRAC camera. The event was initially selected as a ``potentially good" target (i.e., as a ``subjective, secret" event) on 2016 June 12 so that observations would be taken during the first week of the \Spitzer\ campaign, i.e., starting 2016 June 18. We announced the event as a ``subjective, immediate" event on 2016 June 18 at UT 16:34 (HJD$^{\prime} = 7558.19$), just after the start of the \Spitzer\ observations. We discuss the implications of the selection in Section \ref{sec:spzmember}.\\

\begin{deluxetable*}{lr|rrr|lr}
\tablecaption{Best-fit parameters of ground-based models \label{table:ground}}
\tablewidth{0pt}
\tablehead{
\multicolumn{1}{c}{} &
\multicolumn{1}{c}{1L1S} &
\multicolumn{3}{|c}{2L1S} &
\multicolumn{2}{|c}{1L2S} \\
\multicolumn{1}{c}{Parameter} &
\multicolumn{1}{c}{} &
\multicolumn{1}{|c}{STD} &
\multicolumn{1}{c}{APRX ($u_0 < 0$)} &
\multicolumn{1}{c}{APRX ($u_0 > 0$)} &
\multicolumn{1}{|c}{Parameter} &
\multicolumn{1}{c}{}
}
\startdata
$\chi^{2}_{\rm ground} / {\rm N_{data}}$  & $ 11151.00 / 10692   $ & $ 10688.11 / 10692   $ & $ 10684.919 / 10692  $ & $ 10684.597 / 10692  $ & $\chi^{2}_{\rm ground} / {\rm N_{data}}$ & $ 10879.919 / 10692  $ \\  
$t_0$ [${\rm HJD'}$]                      & $ 7560.354 \pm 0.006 $ & $ 7560.317 \pm 0.007 $ & $ 7560.316 \pm 0.007 $ & $ 7560.315 \pm 0.007 $ & $t_{0,{S_1}}$ $[{\rm HJD'}]$             & $ 7560.378 \pm 0.006 $ \\ 
$u_0$                                     & $    0.022 \pm 0.001 $ & $    0.021 \pm 0.001 $ & $   -0.021 \pm 0.001 $ & $    0.021 \pm 0.001 $ & $t_{0,{S_2}}$ $[{\rm HJD'}]$             & $ 7556.494 \pm 0.012 $ \\ 
$t_{\rm E}$ [days]                        & $   54.563 \pm 1.164 $ & $   55.395 \pm 1.214 $ & $   56.222 \pm 1.185 $ & $   56.832 \pm 1.312 $ & $u_{0,{S_1}}$                            & $    0.021 \pm 0.001 $ \\ 
$s$                                       &      \nodata           & $    1.018 \pm 0.001 $ & $    1.017 \pm 0.001 $ & $    1.017 \pm 0.001 $ & $u_{0,{S_2}}$ ($\times10^{-3}$)          & $    0.039 \pm 0.376 $ \\ 
$q$ ($\times10^{-3}$)                     &      \nodata           & $    1.536 \pm 0.104 $ & $    1.463 \pm 0.104 $ & $    1.432 \pm 0.109 $ & $t_{\rm E}$ [days]                       & $   56.815 \pm 1.279 $ \\ 
$\alpha$ [rad]                            &      \nodata           & $    2.528 \pm 0.026 $ & $   -2.537 \pm 0.027 $ & $    2.543 \pm 0.027 $ & $q_{\rm flux}$                           & $    0.003 \pm 0.001 $ \\ 
$\rho_{\ast}$ ($\times10^{-3}$)           &      \nodata           & $    2.123 \pm 0.305 $ & $    2.170 \pm 0.293 $ & $    1.883 \pm 0.297 $ & $\rho_{\ast,{S_1}}$                      &      \nodata           \\ 
$\pieN$                                   &      \nodata           &     \nodata            & $   -0.104 \pm 0.102 $ & $   -0.131 \pm 0.097 $ & $\rho_{\ast,{S_1}}$ ($\times10^{-3}$)    & $    0.921 \pm 0.257 $ \\ 
$\pieE$                                   &      \nodata           &     \nodata            & $    0.256 \pm 0.147 $ & $    0.271 \pm 0.149 $ & \nodata                                  &      \nodata           \\ 
\enddata
\tablecomments{
${\rm HJD' = HJD - 2450000.0}$. For the 1L2S model, the angular radius of the first source ($\rho_{\ast,{S_1}}$) is not measured. 
}
\end{deluxetable*}

{\section{Ground-based Light Curve Analysis}
\label{sec:gb}}

The light curve of \thisevent\ is shown in Figure \ref{fig:lc}. The event increases in brightness by $\sim 3$ magnitudes suggesting a possible high-magnification event. There is a small perturbation at $t_{\rm  pl} = 7556.5$. The best-fit point-source/point-lens (1L1S) model has $t_0 = 7560.3$, $t_{\rm E} = 55$ days, $u_0 = 0.022$, where $t_0$ is the time at the peak of the light curve, $t_{\rm E}$ is the timescale that the source travels the angular Einstein ring radius ($\theta_{\rm E}$), and $u_0$ is the separation in units of $\theta_{\rm E}$ at $t_0$. 

To find the best-fit solution of the binary-lens and single-finite-source (2L1S) model to describe the observed light curve, we first conduct a grid search over the parameters $s$ and $q$. Here, $s$ is the projected separation between binary components, and $q \equiv M_{\rm pl} / M_{\rm h}$ is the mass ratio of binary components, where $M_{\rm pl}$ is the mass of a planet and $M_{\rm h}$ is the mass of a host star. The parameters $s$ and $q$ are related to the caustic morphology. Thus, we set them as grid parameters. For five other parameters of the static 2L1S model (STD), we minimize $\chi^{2}$ using the Markov Chain Monte Carlo algorithm \citep{dunkley05}. These five parameters are $t_0$, $u_0$, $t_{\rm E}$, $\alpha$, and $\rho_{\ast}$. The set of $[t_0, u_{0}, t_{\rm E}]$ are the same as in the 1L1S case, $\alpha$ is the angle of the source trajectory with respect to the binary axis, while $\rho_{\ast}$ is the angular source radius ($\theta_{\ast}$) in units of $\theta_{\rm E}$.

We set a wide range for the $q$ parameter, $\log(q) \in [-5.5, 1.0]$. For the $s$ parameter, the grid range is $\log_{10}(s) \in [-1.0, 1.0]$. Once we find local minima from the grid search, we use a finer grid to search for localized minima. For competitive solutions found from the grid search, we refine models by allowing all parameters to vary. The best-fit solution has $s \sim 1$ and $q \sim 1.5\times10^{-3}$. The detailed parameters of this model are presented as the 2L1S STD solution in Table \ref{table:ground}.

We find that the light curve of \thisevent\ is qualitatively similar to that of KMT-2019-BLG-0842Lb \citep{jung20} or OGLE-2019-BLG-0960Lb \citep{yee21}, both of which have much smaller mass ratios ($q < 10^{-4}$) than our best-fit solution. Following the formalism of \citet{gaudi97}, we estimate the properties of such a solution. Because the planetary perturbation occurs at $u_{\rm pl} \equiv \sqrt{u_{0}^2 + \tau_{\rm pl}^2} = 0.074$, where $\tau_{\rm pl} \equiv (t_{\rm pl} - t_0)/t_{\rm E} = 0.069$, this line of reasoning would imply $s = \left[\sqrt{\left(u_{\rm pl}^2 + 4\right)} + u_{\rm pl}\right]/2 = 1.04$ and $\alpha = \tan^{-1}(u_0/\tau_{\rm pl}) = 162^\circ$ (2.83 radians). Such a trajectory would be very similar to those of  KMT-2019-BLG-0842Lb and OGLE-2019-BLG-0960Lb. Hence, one might expect that \thisevent\ may also contain a planet with a very small mass ratio. Thus, we explicitly search for small mass-ratio solutions similar to KMT-2019-BLG-0842Lb and OGLE-2019-BLG-0960Lb, but find that those solutions are disfavored by $\Delta\chi^{2} \sim 180$ compared to the $(s, q) \sim (1, 1.5\times10^{-3})$ solution.

\subsection{Annual Microlens Parallax (APRX)}

\begin{figure}[htb!]
\epsscale{1.00}
\plotone{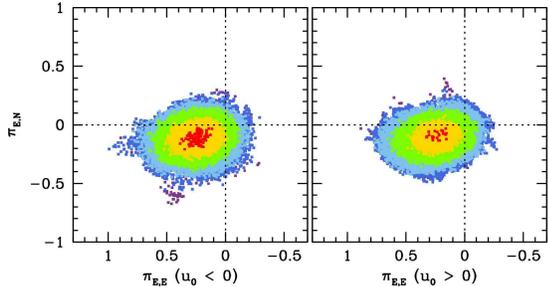}
\caption{
The ($\pieE,\, \pieN$) contours of the APRX models. Each color indicates $\Delta\chi^{2}_{\rm (chain - best)} \leq n^{2}$, 
where $n = 1$ (red), $2$ (yellow), $3$ (green), $4$ (light blue), $5$ (blue), and $6$ (purple), 
respectively. 
\label{fig:APRX}}
\end{figure}

The ground-based light curve analysis shows that the timescale of this event is $\sim 55.4$ days. This is long enough ($t_{\rm E} > 15$ days) to check the effect caused by the annual microlens parallax \citep[APRX;][]{gould92} on the light curve. Thus, we try to check the APRX effect by introducing the microlens-parallax parameters: ($\pieN$, $\pieE$), which indicate the north and east directions of the microlens-parallax vector ($\pivec_{\rm E}$), respectively.

In Table \ref{table:ground}, we present the best-fit parameters of APRX models. From the APRX modeling, we find $\chi^2$ improvements of $\Delta\chi^{2} = 3.2$ and $\Delta\chi^{2} = 3.5$ for ($u_0 < 0$) and ($u_0 > 0$) cases, respectively. These improvements are too minor to claim that the APRX measurements can be used to determine the lens properties. 

However, we find that the APRX contours do not represent a complete non-detection of the APRX effect. The contours are well converged as shown in Figure \ref{fig:APRX}, which gives strong upper limits on the parallax.

\subsection{Lens-orbital Effect}

Because the orbital motion of the binary components can affect the APRX \citep{Batista11}, we also check the lens-orbital (OBT) only model by introducing the lens-orbital parameters: $(ds/dt,\, d\alpha/dt)$ where $ds/dt$ is the variation of the binary separation as a function of time and $d\alpha/dt$ is the variation of the source trajectory angle as a function of time.

We find a negligible $\chi^2$ improvement (i.e., $\Delta\chi^{2}_{\rm (STD-OBT)}=1.67$) when the lens-orbital effect is considered. Also, we find that this very small improvement comes from outside the region of the caustic-crossing. In general, the lens-orbital effect is most sensitive to the caustic-crossing parts of the light curve (that are mostly covered by KMTC and KMTS). Thus, we can conclude that there is no significant lens-orbital effect for this event.

\subsection{2L1S/1L2S Degeneracy}

As \citet{gaudi98} pointed out, the single-lens/binary-source (1L2S) interpretation can mimic planetary anomalies. Also, \citet{shin19} shows that this 2L1S/1L2S degeneracy can appear in wider ranges of cases, especially, the degeneracy appears in cases having non-optimal observational coverage.

For this event, there is a gap in the observational coverage of the second bump. Thus, we check the 2L1S/1L2S degeneracy. For the 1L2S model, we adopt parameters described in \citet{shin19} (A-type, see Appendix of the reference) shown in Table \ref{table:ground}. We find that the $\chi^{2}$ difference between the 1L2S and 2L1S models is $\sim 192$, which is enough to resolve the degeneracy. In particular, the 1L2S cannot properly describe the caustic-crossing feature of the first bump. Thus, we conclude that this event does not suffer from the 2L1S/1L2S degeneracy.

{\section{\Spitzer\ Parallax}
\label{sec:spz_par}}

\subsection{Joint \Spitzer $~+$ Ground}

\begin{figure*}[htb!]
\epsscale{1.10}
\plotone{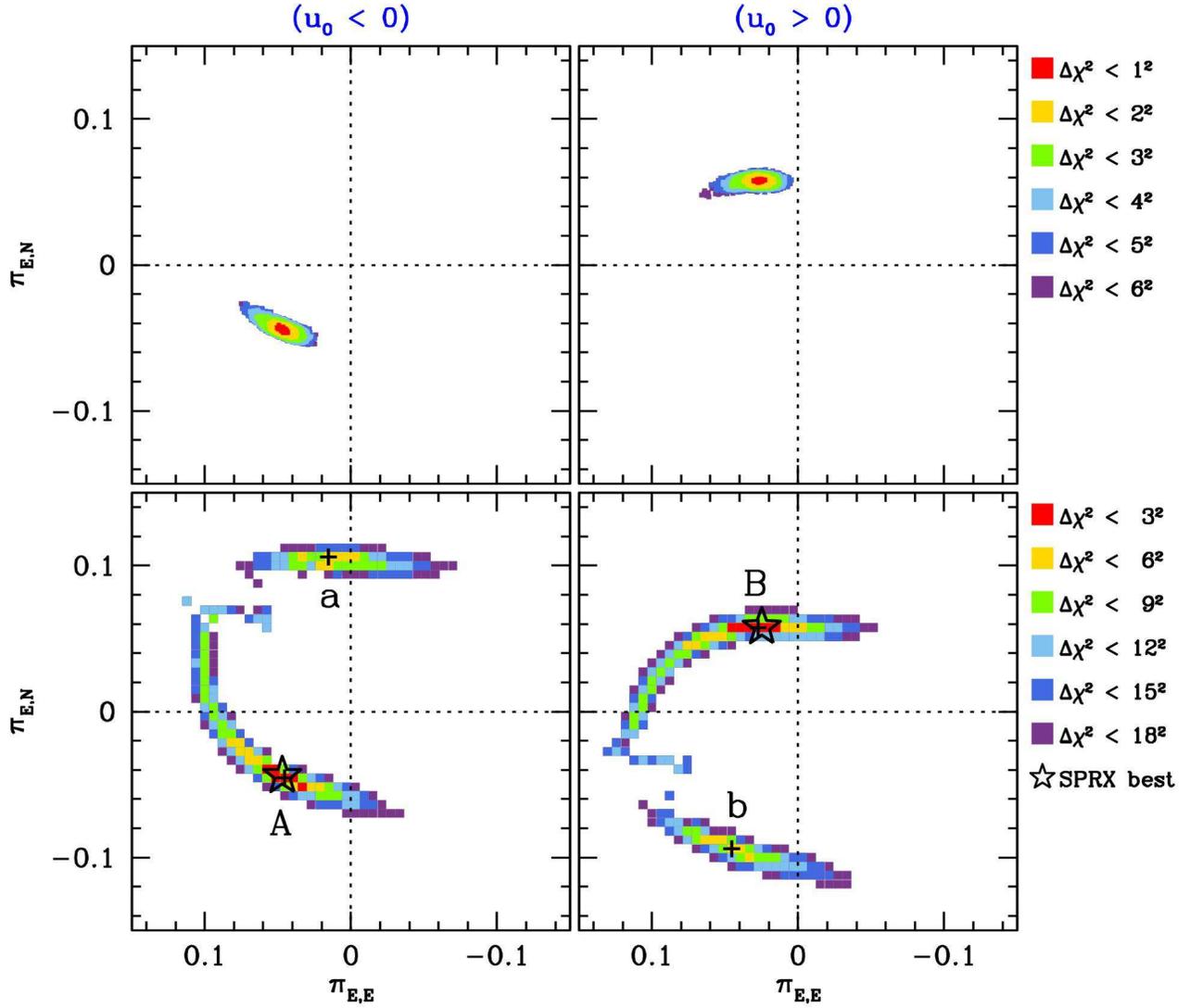}
\caption{
The ($\pieE$, $\pieN$) contours of SPRX models. Top panels show SPRX contours of {\it Joint} 
\Spitzer $~+$ Ground cases. Bottom panels show SPRX contours of \Spitzer-``only" cases. Right and 
left panels present ($u_{0} < 0$) and ($u_{0} > 0$) cases, respectively. The black stars indicate 
the best-fit ($\pieE$, $\pieN$) values of {\it Joint} \Spitzer $+$ Ground cases.
The labeled black $+$'s indicate the ($\pieN$, $\pieN$) locations of the 
four minima shown in Figure \ref{fig:Spz_geo}
\label{fig:SPRX}}
\end{figure*}

To measure the microlens parallax effect using \Spitzer\ data, we jointly fit the \Spitzer\ and ground-based light curves. This fit is constrained by a color-color relation constructed from nearby stars:
\begin{equation}
\label{eqn:cc}
   (I-L)_{\rm pyDIA} = -5.505 \pm 0.025 .
\end{equation}
We note that the color constraint for modeling is $(I-L)_{\rm pySIS} = 2.218 \pm 0.025$ (allowed $2\sigma$ ranges). The value of this color constraint is different from the value shown in Equation \ref{eqn:cc}. Because the \Spitzer\ zero point of the color constraint shown in Equation \ref{eqn:cc} is $25$ magnitude. However, for our modeling scheme, we use an $18$ magnitude as the zero point. Thus, we convert the magnitude system from $25$th to $18$th by adding $7$ magnitudes. In addition, we use the ``pySIS'' photometry of KMTC for modeling to use the best quality of photometry. While, we use the ``pyDIA'' photometry for the source color estimation (the ``pyDIA'' photometry is optimized to obtain $V$-band light curve and the color-magnitude diagrams). These two datasets have different zero points.  For this event, the relation between them is $I_{\rm pySIS} - I_{\rm pyDIA} = 0.723 \pm 0.002$. Thus, the final color constraint for the modeling is that $(I-L)_{\rm pySIS} = (I-L)_{\rm pyDIA} + 7 + 0.723 = 2.218 \pm 0.025$. We apply the color constraint on the models using an additional $\chi^2$ (i.e., $\chi^{2}_{\rm penalty}$) that is weighted by the different amount of the source color. The details of $\chi^{2}_{\rm penalty}$ are described in \citet{shin17}.

The top panels of Figure \ref{fig:SPRX} show that the resulting constraints on the parallax are very tight for the both $(u_0 < 0)$ and $(u_0 > 0)$ solutions. These panels show two well-constrained minima at $\pivec_{\rm E}(N, E) = (-0.044, 0.047)$ and $(0.058, 0.025)$ for the $(u_0 < 0)$ and $(u_0 > 0)$ solutions, respectively. 
This is in contrast to the ``usual" situation for point lens events, which typically show a four-fold degeneracy \citep{refsdal66, gould94} or an arc \citep[often seen for partial light-curves like this one;][]{gould19}. We revisit this degeneracy in the next section. These minima imply parallax values of $\pi_{\rm E} = 0.064$ and $0.063$, respectively. They are also consistent with the broad constraints on the parallax from the annual parallax effect (see Figure \ref{fig:APRX}).  The full fits are given in Table \ref{table:space}.

\begin{figure*}[htb!]
\epsscale{1.10}
\plotone{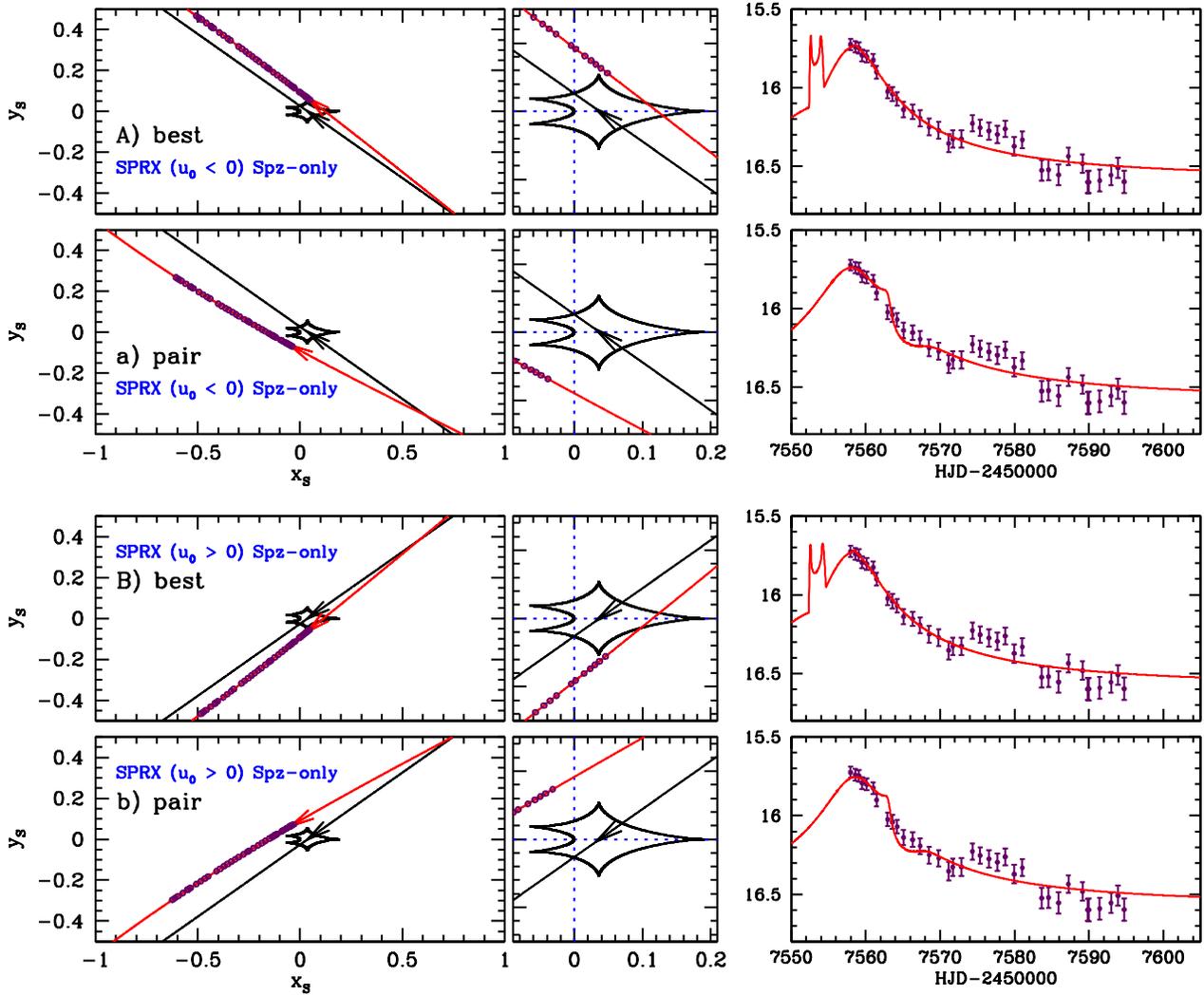}
\caption{
Four solutions from \Spitzer-``only" fitting. Left panels show the trajectories of the source as seen 
from the ground (black line with arrow) and \Spitzer\ (red line with arrows) over the full span of 
the observations from \Spitzer\ (epochs of observations shown as purple circles). The caustics are 
shown as a black, closed curve. Center panels show a zoom of the caustic region. Right panels show 
the \Spitzer\ light curve (purple points) with the best-fit model light curve (red line) for each 
solution. The ``a" and ``b" solutions are disfavored relative to the ``A" and ``B" solutions; 
lettering corresponds to the minima indicated in Figure \ref{fig:SPRX}.
\label{fig:Spz_geo}}
\end{figure*}

\begin{deluxetable*}{l|rr|rr}
\tablecaption{Best-fit parameters of SPRX Tests with the ($I-L$) color constraint \label{table:space}}
\tablewidth{0pt}
\tablehead{
\multicolumn{1}{c}{} &
\multicolumn{2}{|c|}{SPRX} &
\multicolumn{2}{c}{Cheap-SPRX} \\ 
\multicolumn{1}{c}{Parameter} &
\multicolumn{1}{|c}{($u_0 < 0$)} &
\multicolumn{1}{c}{($u_0 > 0$)} &
\multicolumn{1}{|c}{($u_0 < 0$)} &
\multicolumn{1}{c}{($u_0 > 0$)}
}
\startdata
$\chi^{2}_{\rm Total}$                    & $ 10735.013          $ & $ 10737.220          $ & $ 10687.513                  $ & $ 10686.936                  $ \\  
$\chi^{2}_{\rm ground}  / {\rm N_{data}}$ & $ 10686.57 / 10692   $ & $ 10689.34 / 10692   $ & $ 10686.55 / 10692           $ & $ 10686.07 / 10692           $ \\  
$\chi^{2}_{\rm Spitzer} / {\rm N_{data}}$ & $ 48.28 / 36         $ & $ 47.70 / 36         $ & $ 0.84 / 4                   $ & $ 0.84 / 4                   $ \\ 
$\chi^{2}_{\rm penalty}$                  & $ 0.164              $ & $ 0.125              $ & $ 0.121                      $ & $ 0.030                      $ \\ 
$t_0$ [${\rm HJD'}$]                      & $ 7560.318 \pm 0.007 $ & $ 7560.319 \pm 0.007 $ & $ 7560.315 \pm 0.007         $ & $ 7560.318 \pm 0.007         $ \\ 
$u_0$                                     & $   -0.021 \pm 0.001 $ & $    0.021 \pm 0.001 $ & $   -0.021 \pm 0.001         $ & $    0.021 \pm 0.001         $ \\ 
$t_{\rm E}$ [days]                        & $   56.355 \pm 1.168 $ & $   56.311 \pm 1.242 $ & $   55.401 \pm 1.241         $ & $   56.370 \pm 1.241         $ \\ 
$s$                                       & $    1.017 \pm 0.001 $ & $    1.017 \pm 0.001 $ & $    1.018 \pm 0.001         $ & $    1.018 \pm 0.001         $ \\ 
$q$ ($\times10^{-3}$)                     & $    1.473 \pm 0.099 $ & $    1.470 \pm 0.102 $ & $    1.523 \pm 0.105         $ & $    1.469 \pm 0.104         $ \\ 
$\alpha$ [rad]                            & $   -2.532 \pm 0.026 $ & $    2.535 \pm 0.025 $ & $   -2.532 \pm 0.026         $ & $    2.531 \pm 0.026         $ \\ 
$\rho_{\ast}$ ($\times10^{-3}$)           & $    2.142 \pm 0.303 $ & $    2.103 \pm 0.309 $ & $    2.216 \pm 0.312         $ & $    2.185 \pm 0.312         $ \\ 
$\pieN$                                   & $   -0.044 \pm 0.003 $ & $    0.058 \pm 0.002 $ &      \nodata                   &      \nodata                   \\ 
$\pieE$                                   & $    0.047 \pm 0.006 $ & $    0.025 \pm 0.006 $ &      \nodata                   &      \nodata                   \\ 
$\Apie$                                   & $    0.064 \pm 0.003 $ & $    0.063 \pm 0.003 $ & $    0.066_{-0.010}^{+0.013} $ & $    0.088_{-0.027}^{+0.007} $ \\ 
$f_{\rm S, KMTC\, {\it(I)}}$              & $    0.044 \pm 0.001 $ & $    0.044 \pm 0.001 $ & $    0.045 \pm 0.001         $ & $    0.044 \pm 0.001         $ \\ 
$f_{\rm S, Spitzer, {\it(L)}}$            & $    0.341 \pm 0.011 $ & $    0.341 \pm 0.011 $ & $    0.341 \pm 0.011         $ & $    0.339 \pm 0.011         $ \\ 
$(I-L)_{\rm pySIS}$                       & $    2.223 \pm 0.043 $ & $    2.223 \pm 0.043 $ & $    2.209 \pm 0.043         $ & $    2.223 \pm 0.043         $ \\ 
\enddata
\end{deluxetable*}

{\subsection{\Spitzer-``only"}
\label{sec:sponly}}

As a check, we also conduct the \Spitzer-``only" test for the full \Spitzer\ dataset. We conduct this  \Spitzer-``only" analysis following the formalism laid out in \citet{Gould20_0029} but using a full planet model as in \citet{zang20}. Specifically, we fix the seven parameters of the model ($t_0$, $u_0$, $t_E$, $\rho$, $\alpha$, $s$, $q$) to their ground-based values, vary $\pi_{E, N}$ and $\pi_{E, E}$ on a grid, and fit only the \Spitzer\ data with the color-constraint applied. The resulting parallax constraints shown in the bottom panels of Figure \ref{fig:SPRX} are more extended arcs compared to the constraints from the full, joint fit. In fact, they show four local minima as might be expected from the four-fold degeneracy discussed above. Nevertheless, the four-fold degeneracy is broken because the caustic structure induces additional structure to the light curve (see Figure \ref{fig:Spz_geo}), and the \Spitzer-``only" fits ultimately recover the same global minimum in each case ($u_{0} > 0$ and $u_{0} < 0$). Then, comparing the top and bottom panels in Figure \ref{fig:SPRX} shows that the joint fitting further suppresses these alternate minima leaving a two-fold degeneracy rather than a four-fold degeneracy.\\

{\section{Tests of the \Spitzer\ Parallax}
\label{sec:Spz_data}}

\begin{figure}[htb!]
\epsscale{1.00}
\plotone{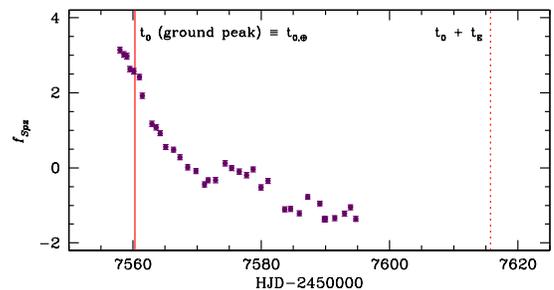}
\caption{
{\it Spitzer} light curve in instrumental flux units. 
The solid vertical line indicates the peak of the light curve as observed from Earth 
($t_{0} \equiv t_{0,\oplus}$) and the dotted vertical line is $+t_{\rm E}$ away.
\label{fig:spitzer_flux}}
\end{figure}

\citet{KoshimotoBennett19} have suggested that systematics in the \Spitzer\ light curve may bias the resulting parallax measurements. Such systematics have been seen at the level of 1--2 instrumental flux units \citep{Gould20_0029, hirao20, zang20}, and so are most likely to play a significant role in events with small changes in flux as measured by \Spitzer. Figure \ref{fig:spitzer_flux} shows the \Spitzer\ light curve of \thisevent\ in instrumental flux units and illustrates that it is in this regime. However, because \thisevent\ is in the high-magnification regime, this permits several tests that can be used to verify the parallax signal, which we describe below.
\\

\subsection{Heuristic Analysis}
\label{sec:heuristic}

First, overall, the \Spitzer\ light curve appears to decline during the \Spitzer\ observation window, suggesting that the event peaked earlier as seen from the ground. Because \Spitzer\ was separated from Earth by $D_{\perp} = 1.021$ au (as projected on the sky), this suggests $\pi_{\rm E} > \Delta \tau\, (\au / D_{\perp})$ where $\tau = (t_{0, \oplus} - t_{\rm start, Spz}) / t_{\rm E} = 0.043$, so $\pi_{\rm E} > 0.042$, which is consistent with the values derived from the full-fit.

Second, the characteristics of \thisevent\ are similar to the criteria set out by \citet{gould12} for ``cheap-space parallaxes". Specifically, the event is in the high-magnification regime ($A_{\rm peak} \sim 50$), the \Spitzer\ observations start before the ground-based peak ($t_{\rm start, Spz} = 7558.0 < t_{0,\oplus}  = 7560.3$), and, as we show below, was observed by \Spitzer\ close to baseline. The basic logic is that if \Spitzer\ observes at the peak as seen from the ground, then $u_{\oplus} \sim 0$ and
\begin{equation}
	\pi_{\rm E} \sim \frac{\au}{D_{\perp}}u_{\rm Spz}.
\end{equation}

In this case, we observe $\Delta f_{\rm Spz} \sim 4$ flux units (between $t_{0,\oplus}$ and $t_{\rm last, Spz}$), and we can estimate $f_{\rm source, Spz}$ from the color-color relation in Equation \ref{eqn:cc}. 
Given $I_{\rm source} = 20.9$, this implies $f_{\rm source, Spz} = 0.274$.
The last \Spitzer\ observation is taken at HJD$^{\prime} = 7589.1$ when the event is magnified by $A_{\oplus} = 2.1$ as seen from Earth. Because we know the event peaked earlier from \Spitzer, we know that $A_{\rm last, Spz} < 2.1$. Then,
\begin{equation}
\begin{split}
	\Delta f_{\rm Spz} & = f_{\rm Spz}(t_{0,\oplus}) - f_{\rm Spz}(7589.1) \\
                           & = f_{\rm source, Spz} [A_{\rm Spz}(t_{0,\oplus}) - A_{\rm last, Spz}],
\end{split}
\end{equation}
Hence, $15.6 < A_{\rm Spz}(t_{0, \oplus}) < 16.7$. This yields $0.060 < u_{\rm Spz}(t_{0, \oplus}) < 0.064$ and an estimate of the parallax: $0.059 < \pi_{\rm E} < 0.063$ (for $D_{\perp} = 1.021$ au), although the parallax could be somewhat larger for $u_{\oplus} > 0$. 

An additional caveat is that the ``cheap-space parallaxes" formalism was developed for point lenses, whereas \thisevent\ contains a large resonant caustic structure. This caustic structure has the potential to cause this formalism to break down. At the same time, we can already see that the measured values of the parallax are in good agreement with this simplified heuristic analysis.

\begin{figure*}[htb!]
\epsscale{1.00}
\plotone{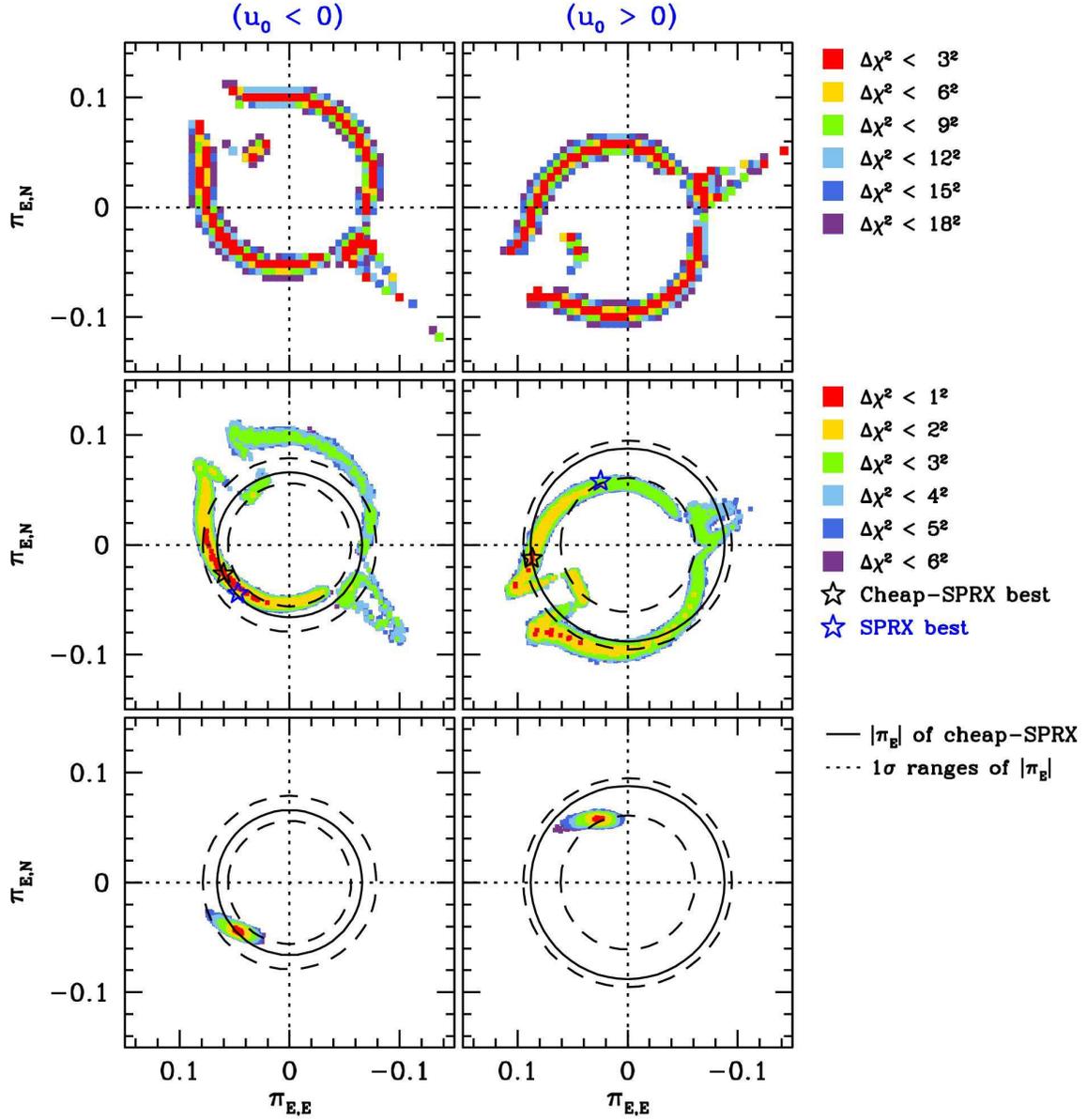}
\caption{
The ($\pieE$, $\pieN$) contours of cheap--SPRX models with SPRX contours for comparison. Top panels 
show cheap--SPRX contours of \Spitzer--``only" cases. Middle panels show cheap--SPRX contours of 
{\it Joint} \Spitzer $~+$ Ground cases. The bottom panels show the SPRX contours for comparison. 
The right and left panels present the ($u_{0} < 0$) and ($u_{0} > 0$)cases, respectively. The circles 
indicate $\Apie$ values (solid lines) measured from the cheap--SPRX with their $1\sigma$ errors 
(dashed lines). The blue and black star marks indicate the best-fit values of SPRX and cheap--SPRX 
models, respectively. 
\label{fig:cheapSPRX}}
\end{figure*}

\subsection{Cheap-Space Parallax Limit}
\label{sec:cheapsprx}

To further explore the application of ``cheap-space parallaxes" to this event, we next fit a subset of the \Spitzer\ data, as might be obtained by such an observational program.
Specifically, we restrict the fitting to the \Spitzer\ observation taken closest to $t_{0,\oplus}$ and the last three \Spitzer\ observations (technically, only the last observation is needed, but the observed scatter in the light curve suggests that, in this case, a single observation may not accurately reflect the mean magnification). Under the ``cheap-space parallaxes" formalism, for a points lens with $u_0 = 0$, we would expect this test to produce an annulus on the $\pivec_{\rm E}$ plane centered at the origin \citep[e.g.,][]{shin18}.

\begin{figure*}[htb!]
\epsscale{1.00}
\plotone{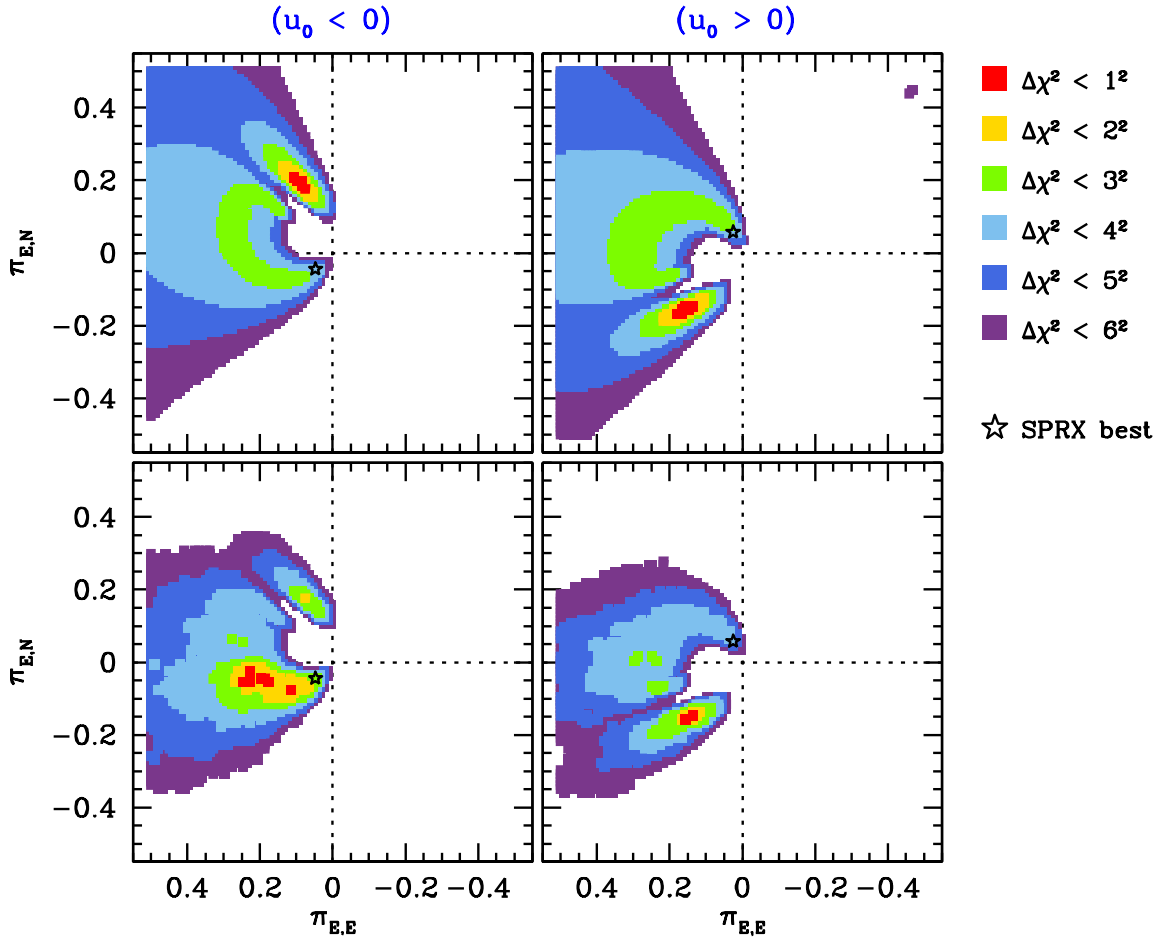}
\caption{
Constraints on the microlens parallax vector {\it without} including the color constraint. 
The top panels show \Spitzer-``only" cases. The bottom panels show \Spitzer $+$ Ground cases. 
The star mark indicates the best-fit parallax {\it with} including the color constraint.
These differ significantly from the constraints in Figure \ref{fig:SPRX} because the correlated noise 
in the \Spitzer\ light curve can be fit by the planetary model if an arbitrary source flux is allowed.
\label{fig:without_CC}}
\end{figure*}

\subsubsection{\Spitzer-``only" Cheap-Space Parallaxes}

For the first test, we conduct \Spitzer-``only" fitting with this restricted dataset.
The full contours are shown in the top panels of Figure \ref{fig:cheapSPRX}. They show that, for this event with its large resonant caustic, the contours are still similar to a ring shape. That ring is displaced from the origin by $\sim u_{0, \oplus}$, as would be expected for the 1L1S case with $|u_{0, \oplus}| > 0$. The uncertainty in $\pi_{\rm E}$ due to $|u_{0, \oplus}| > 0$ is comparable to the uncertainty due to 2L1S departures from the ring-shape. This indicates that the ``cheap-space parallaxes" formalism still applies in 2L1S cases with $q \ll 1$ (i.e., the magnification map is still dominated by the effect of the primary lens). 

Finally, from this fit, we derive $0.06 < \Apie < 0.10$, which much stronger than the constraint from APRX alone and constrains the lens to be in the bulge (when combined with $\theta_{\rm E}$, see Section \ref{sec:cmd}). 

\subsubsection{\Spitzer\ $+$ Ground Cheap-Space Parallaxes}

We also conduct a joint fit to the ground-based data and the ``cheap-space parallax" subset of the \Spitzer\ data (and including the color-constraint). The results of the joint fit are shown in the middle panels of Figure \ref{fig:cheapSPRX} and the parameters are given in Table \ref{table:space}. These contours show the influence of the APRX constraint from the ground-based data, which weakly prefer parallaxes in the middle-left panel. Hence, in the ``cheap-space parallax" limit, even weak constraints or upper limits from APRX can be meaningful.

\subsection{Test of \Spitzer\ Systematics}

\subsubsection{Role of the Color-constraint}

Because the overall change in the \Spitzer\ flux is small relative to the magnitude of the systematics, the color-constraint plays a very important role in this event. When we fit the \Spitzer\ data without including the color-constraint, we find a completely different solution for the parallax. Figure \ref{fig:without_CC} shows the contours for the \Spitzer-``only" case {\it without} the color-constraint (the joint fit is similar because of the weakness of the APRX). For the best-fit solution, $f_{\rm S,Spz} \sim 0.9$, which a factor of $\sim 3$ larger than the value expected from the color-constraint (Section \ref{sec:heuristic}). Close inspection of the \Spitzer\ light curve in Figure \ref{fig:spitzer_flux} shows an apparent ``dip" in the light curve between HJD$^{\prime} \sim 7561$ and HJD$^{\prime} \sim 7574$ at the level of $\sim 1$ flux unit compared to a smooth decline. Thus, in the absence of the color-constraint, this dip can be fit by the ``trough" induced by the $s < 1.02$ planet. This confirms that correlated noise can exist in the \Spitzer\ light curves at the level of $\sim 1$ flux unit.

\begin{figure}[htb!]
\epsscale{1.10}
\plotone{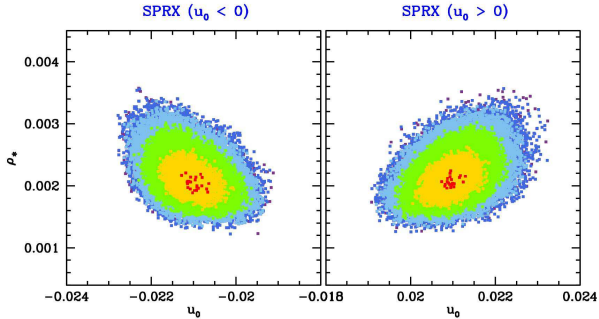}
\caption{
The contours of the $\rho_{\ast}$ parameter. The color scheme is identical to Figure \ref{fig:APRX}.
\label{fig:rho}}
\end{figure}

\subsubsection{Potential Impact of Systematics}

Systematics at the level of $\sim 1$ flux unit have been seen in several events \citep[e.g.,][]{Poleski16,Dang20,Gould20_0029,hirao20} and are confirmed in this case. The potential impact of systematics will be most pronounced in events for which the overall flux change is small (e.g., for $\Delta f_{\rm Spz} \lesssim 10$ flux units). For example, such an impact was seen in KB180029 \citep{Gould20_0029}, for which the baseline flux differed by $\sim 1$ flux unit between seasons.

Because the overall flux change in \thisevent\ is only $\Delta f_{\rm Spz} \sim 4$ flux units, we briefly consider how such systematics might affect the parallax. We have already shown that the ``cheap-space parallaxes" formalism is a reasonable approximation in this case and that our simple estimate of the parallax from Section \ref{sec:heuristic} is reasonably accurate. If systematics affect either (or both) the peak or the baseline of the \Spitzer\ light curve for \thisevent, then the true change in flux might be as low as 3 flux units or as high as 5 flux units. Taking into account this range of fluxes (and the uncertainty in $A_{\rm last}$), the parallax would still be confined to the range $\pi_{\rm E} = [0.05, 0.08]$, yielding a factor of $\sim 1.3$ uncertainty in the lens mass and a $\sim 0.25$ kpc uncertainty in the distance. Hence, the lens is still constrained to be a low-mass dwarf in the bulge.

{\section{Source Color and Lens Properties}
\label{sec:cmd}}

\begin{figure}[htb!]
\epsscale{1.10}
\plotone{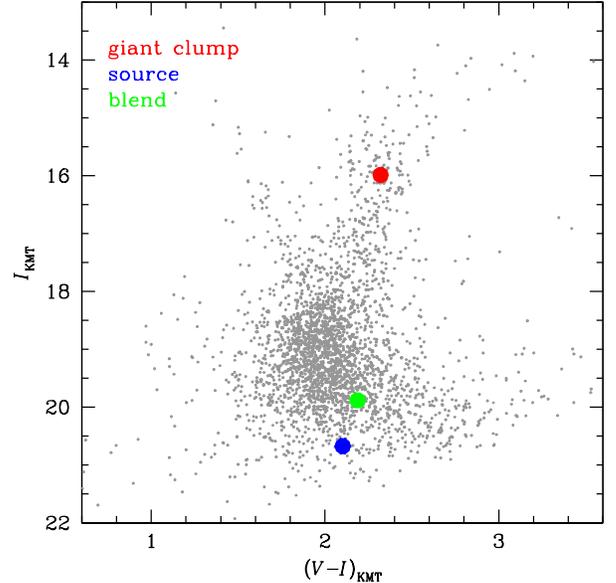}
\caption{
Color-magnitude diagrams of the KMTNet. 
\label{fig:CMD}}
\end{figure}

\subsection{Finite-source Effect}

In addition to measuring $\pi_{\rm E}$, one must also determine the angular Einstein ring radius ($\theta_{\rm E}$), in order to derive the lens properties such as mass of the lens system ($M_{\rm L}$ and the distance to the lens system ($D_{\rm L}$): 
\begin{equation}
M_{\rm L} = \frac{\theta_{\rm E}}{\kappa \pi_{\rm E}}, ~~ 
D_{\rm L} = \frac{\rm au}{\pi_{\rm E} \theta_{\rm E} + \pi_{\rm S}},
\label{eqn:lens}
\end{equation}
where $\kappa = 8.144\, {\rm mas}\, M_{\odot}^{-1}$, $\pi_{\rm S}\equiv {\rm au}/D_{\rm S}$ is the parallax of the source, and $D_{\rm S}$ is a distance to the source. 

The $\theta_{\rm E}$ can be determined using the finite-source effect: $\rho_{\ast} \equiv \theta_{\ast}/\theta_{\rm E}$ where $\theta_{\ast}$ is the angular source radius, which is an observable that can be routinely measured (see Section \ref{sec:source_color}). For this event, there are caustic-crossing features that are well-covered by KMTS and KMTC. Thus, it is possible to measure $\rho_{\ast}$ from the finite-source effect. In Figure \ref{fig:rho}, we present the contours of $\rho_{\ast}$. The contours show that the $\rho_{\ast}$ values are securely measured.

\begin{deluxetable*}{l|rr|rr}
\tablecaption{Lens properties determined using SPRX and cheap--SPRX \label{table:properties}}
\tablewidth{0pt}
\tablehead{
\multicolumn{1}{c}{} &
\multicolumn{2}{|c}{SPRX} &
\multicolumn{2}{|c}{Cheap-SPRX} \\ 
\multicolumn{1}{c}{Properties}  &
\multicolumn{1}{|c}{($u_0 < 0$)} &
\multicolumn{1}{c}{($u_0 > 0$)} &
\multicolumn{1}{|c}{($u_0 < 0$)} &
\multicolumn{1}{c}{($u_0 > 0$)}
}
\startdata
$\theta_{\rm E}$ (mas)            & $ 0.24 \pm 0.04 $ & $ 0.25 \pm 0.04 $ & $ 0.23 \pm 0.04 $ & $ 0.24 \pm 0.04 $ \\ 
$M_{\rm host}$ ($M_{\odot}$)      & $ 0.46 \pm 0.08 $ & $ 0.48 \pm 0.09 $ & $ [0.31, 0.60]  $ & $ [0.25, 0.56]  $ \\
$M_{\rm planet}$ ($M_{\rm Jup}$)  & $ 0.71 \pm 0.12 $ & $ 0.74 \pm 0.13 $ & $ [0.49, 0.96]  $ & $ [0.39, 0.86]  $ \\
$D_{\rm S}$ (kpc)                 & $ 9.40 \pm 0.87 $ & $ 9.39 \pm 1.21 $ & $ 9.16 \pm 1.08 $ & $ 9.20 \pm 1.09 $ \\
$D_{\rm L}$ (kpc)                 & $ 8.12 \pm 0.69 $ & $ 8.06 \pm 0.91 $ & $ 7.93 \pm 0.84 $ & $ 7.79 \pm 0.82 $ \\ 
$D_{\rm LS}$ (kpc)                & $ 1.28 \pm 0.27 $ & $ 1.33 \pm 0.37 $ & $ 1.23 \pm 0.36 $ & $ 1.41 \pm 0.43 $ \\
$a_{\perp}$ (au)                  & $ 2.13 \pm 0.33 $ & $ 2.23 \pm 0.43 $ & $ 2.01 \pm 0.34 $ & $ 2.00 \pm 0.34 $ \\
\enddata
\end{deluxetable*}

{\subsection{Angular Source Radius}
\label{sec:source_color}}

The angular source radius can be measured using the conventional method \citep{yoo04}. The method requires multi-band observations. The KMTNet survey regularly observes $V$-band data. In $2016$, KMTC made one $V$-band observation for every $10$ $I$-band observations, while KMTS made $V$-band observations at half this rate. We use the KMTC observations to determine the color. Using the multi-band observations, we construct the KMTNet color-magnitude diagrams (CMD) shown in Figure \ref{fig:CMD}. Then, we measure the reddened and de-reddened colors of the source using Yoo et al.'s method and the intrinsic color and magnitude of the red giant clump adopted from \citet{bensby11} and \citet{nataf13}:
\begin{equation}
(V-I) = 2.102 \pm 0.013,
\end{equation}
\begin{equation}
(V-I)_{0} = 0.842 \pm 0.052.
\end{equation}

We estimate the angular source radius using the color conversion of \citet{BB88} and the color/surface-brightness relation of \citet{kervella04}:
\begin{equation}
\theta_{\ast} = 0.52 \pm 0.03\, \mu{\rm as}
\end{equation}
Then, by combining the $\rho_{\ast}$ measurements, we determine the angular Einstein ring radius of each solution shown in Table \ref{table:properties}.

{\subsection{Lens properties}
\label{sec:lens_properties}}

In Table \ref{table:properties}, we present lens properties determined using measurements of $\Apie$ and $\theta_{\rm E}$ (see Equation \ref{eqn:lens}). The binary lens system is revealed as a planetary system consisting of a sub-Jupiter-mass planet ($M_{\rm pl} \sim 0.7\, M_{\rm Jup} $) orbiting an M-dwarf host star ($M_{\rm h} \sim 0.5\, M_{\odot}$) with a projected separation of $\sim 2$ au. 

Because the source distance is not precisely known, $\pi_{\rm rel}$ (and therefore $D_{\rm LS} \equiv D_{\rm S}-D_{\rm L} \simeq \left( \pi_{\rm rel}/{\rm au} \right) D_{\rm clump}^2$) is much better constrained than $D_{\rm L}$. Here, we adopt $D_{\rm clump} = 8.54$ kpc from \citet{nataf13}. We find 
\begin{equation}
D_{\rm LS} \simeq \frac{\pi_{\rm rel}}{\rm au} D_{\rm clump}^2 = 1.1 \pm 0.2\, {\rm kpc}
\label{eqn:pirel}
\end{equation}
Because, the source is almost certainly in the bulge, this small value of $D_{\rm LS}$ provides strong evidence that the lens is in the bulge as well.

Moreover, we estimate distances to lens ($D_{\rm L}$) and source ($D_{\rm S}$) using the Bayesian analysis with constraints of $t_{\rm E}$, $\theta_{\rm E}$, and $\pivec_{\rm E}$. The Bayesian formalism is adopted from \citet{shin21} with the mass function of \citet{chabrier03}. The Bayesian results indicate that the lens is located at the bulge ($D_{\rm L} \sim 8.1$ kpc). This result directly supports the $D_{\rm LS}$ argument, which shows the planet is located in the bulge. We present the distances of each case in Table \ref{table:properties}.
\\

{\section{Membership in the \Spitzer\ sample}
\label{sec:spzmember}}

According to the protocols of \citet{yee15}, because the first \Spitzer\ observation (at HJD$^{\prime} = 7557.96$) was taken before the event was announced as a \Spitzer\ target, this point must be excluded from the evaluation of the \Spitzer\ parallax for the purposes of evaluating whether or not the event is part of the statistical \Spitzer\ planet sample\footnote{However, if the event is found to be in the sample based on the limited dataset, the full dataset can be used to {\it characterize} the parallax.}. Therefore, we repeat the joint ground+\Spitzer\ fitting but without the first \Spitzer\ observation. We find $\Apie = 0.062 \pm 0.003$ and $\Apie = 0.063 \pm 0.003$ for the $(u_0 < 0)$ and $(u_0 > 0)$ solutions, respectively.

In terms of calculating membership in the \Spitzer\ sample, we need to calculate $D_{8.3}$ and its uncertainty (i.e., the lens distance for a source at 8.3 kpc, see \citealt{zhu17} for details). We find $D_{8.3} = 7.3$ kpc and $\sigma_{D_{8.3}} = 0.2$ kpc. According to \citet{zhu17}, events with $\sigma_{D_{8.3}} < 1.4$ kpc can be included in the statistical \Spitzer\ sample. Hence, \thisevent\ meets this criterion.

Then, we should consider whether or not the planet in \thisevent\ can be included in the sample. \citet{yee15} have specified that only planets (and planet sensitivity) from after the selection can be included in the statistical analysis. In this case, the event was not selected until HJD$^{\prime}=7558.19$, which is after the planet perturbation at $\sim 7556.5$. However, the last datum that was available to the \Spitzer\ team when it made its decision was at $7555.63$, i.e., before the anomaly. That is, although at the time of the decision, OGLE had taken one additional observation (at $7557.57$), this was not posted to the OGLE web page until $7558.29$, i.e., after the decision.  Moreover, MOA did not issue its alert until $7564.05$, and KMT did not reduce its data until after the season. The anomaly was first recognized by KMTNet member K.-H. Hwang in May 2019. Hence, no information about the planet (or possible planet) was available to the team at the time of the decision and all of the KMT and MOA data can be included in the statistical analysis.
\\

{\section{Conclusions}
\label{sec:conclu}}

Using \Spitzer\, parallax combined with finite source effects, we find that OGLE-2016-BLG-1093Lb is a sub-Jupiter-mass planet with the mass of $0.59$--$0.87\, M_{\rm Jup}$. This planet lies beyond the snow line ($a_{\perp} \sim 2$ au) of an M-dwarf host with the mass of $0.38$--$0.57\, M_{\odot}$ \citep[$a_{\rm snow} = 1.0$--$1.5$ au:][]{kennedy08}. This planet is part of the statistical sample of \Spitzer\ microlensing planets with measured distances. Although the planet perturbation occurred prior to the selection of the event for \Spitzer\ observations, no anomalous data points were available to the \Spitzer\ Team until after it made its selection, and so it meets the criteria of \citet{yee15} for inclusion in the sample. Including, \thisevent Lb, the total number of planets in the sample is now eight (two-thirds of the expected number for the program). Moreover, the recent discovery of the planet in OGLE-2019-BLG-1053 \citep{Zang21_AF1} suggests additional planets remain undiscovered in the \Spitzer\ sample.

Because this is a high magnification event ($A_{\rm max} \sim 50$), we are able to conduct a number of tests of the SPRX measurement. First, we conducted a test by adopting the cheap--SPRX idea from \citet{gould12}, in which we use only the \Spitzer\ data taken closest to the ground-based peak of the event and the least magnified points. From this test, we found that the cheap--SPRX measurement is consistent with the full SPRX measurement at the $1\sigma$ level. This test demonstrates that the cheap space-parallax method can be applied to 2-body lenses in addition to point lenses, at least in cases for which the 2-body perturbation is small relative to the overall magnification effect.

We also perform the \Spitzer-``only" test from \citet{Gould20_0029} and investigate systematics in the \Spitzer\ light curve. Because the annual parallax signal is very weak, the \Spitzer-``only" test is very similar to the joint \Spitzer+ ground fitting for the full dataset; i.e., the parallax measurement is dominated by the SPRX. However, it has some influence on the ``cheap--SPRX" fits because relatively little \Spitzer\ data are used. Finally, we confirm that there are systematics in the \Spitzer\ light curve of \thisevent\ at the $\sim 1$ flux unit level. In a completely free fit, these systematics may be fit by features in the planetary magnification pattern. However, such models are ruled out once the constraint on the source flux is included.

In addition to \thisevent Lb, there are several other cases of giant planets with distances measured to be in or very near the bulge. Based on measurements of the lens flux resolved from the source, \citet{Vandorou20} find $m_p = 2.74\, M_{\rm Jup}$, $D_L = 6.72$ kpc for MOA-2013-BLG-220Lb, and \citet{Bhattacharya21} find $m_p = 1.71\, M_{\rm Jup}$, $D_L = 6.89$ kpc for MOA-2007-BLG-400Lb. \citet{Batista14} measured a flux excess for MOA-2011-BLG-293 and inferred $m_p = 4.8\, M_{\rm Jup}$, $D_L = 7.72$ kpc assuming this excess is due to the lens. However, \citet{Koshimoto20} have suggested this excess is due to a source companion rather than lens; nevertheless, they still infer a giant planet in or near the bulge. In addition, the planet in OGLE-2016-BLG-1093 is similar to OGLE-2017-BLG-1140Lb, which is $1.6\, M_{\rm Jup}$ planet orbiting a $0.2\, M_\odot$ host with $D_{8.3} = 7.4\, $ kpc \citep[see ][for the definition of $D_{8.3}$]{zhu17}; although, in that case, the measured proper motions suggested that the lens was a member of the disk population. Regardless, the growing number of detected giant planets at bulge distances would seem to contradict the expectation from \citet{thompson13} that giant planets cannot form in the bulge. \thisevent\ also suggests a definitive answer will be possible once analysis of the full \Spitzer\ sample is complete.

\hbox{}
{\it Acknowledgments.}
This research has made use of the KMTNet system operated by the Korea Astronomy and Space Science Institute (KASI), and the data were obtained at three host sites of CTIO in Chile, SAAO in South Africa, and SSO in Australia.  
Work by C.H. was supported by grants of the National Research Foundation of Korea (2017R1A4A1015178 and 2019R1A2C2085965).
J.C.Y. acknowledges support from N.S.F Grant No. AST--2108414 and JPL grant 1571564.
The MOA project is supported by JSPS KAK--ENHI Grant Number JSPS24253004, JSPS26247023, JSPS23340064, JSPS15H00781, JP16H06287,17H02871 and 19KK0082.


\end{document}